\documentclass{article}

\usepackage{graphicx}
\usepackage{amsmath}
\usepackage{amssymb}
\usepackage{epsfig}
\usepackage{mhchem}
\usepackage{color}

\title{A new open source software for the calculation of the liquid junction
potential between two solutions according to the stationary Nernst--Planck equation}
\author{M. Marino, L. Misuri, D. Brogioli}

\begin{document}

\maketitle

\begin{abstract}
We describe an open source software which we have realized and
made publicly available at the website
http://jljp.sourceforge.net. It provides the potential difference
and the ion fluxes across a liquid junction between the solutions
of two arbitrary electrolytes. The calculation is made by solving
the Nernst--Planck equations for the stationary state in
conditions of local electrical quasi-neutrality at all points of
the junction. The user can arbitrarily assign the concentrations
of the ions in the two solutions, and also specify the analytical
dependence of the diffusion coefficient of each ion on its
concentration.

\end{abstract}

\section{Purpose and description of the software}

A liquid junction potential develops when two solutions containing
ions of different species and/or concentrations come into contact
\cite{macinnes}. Although it may affect in a significant way
several types of electrochemical measurements, it is very
difficult to measure directly in a given experimental setup. Hence
a proper correction of the results for the presence of liquid
junction potentials requires a tool for their theoretical
calculation. Commercial programs exist, specifically designed for
biomedical applications \cite{barry}, with which such a
calculation can be performed, but to our knowledge until now there
existed no open source software freely available to the scientific
community, and easy to use in the most general situations. The
program we are presenting in this paper is written in Java and is
available as an applet, with a simple and user-friendly graphical
panel, at the site http://jljp.sourceforge.net. In order for the
program to be run via a web browser, the Java plugin must have
been installed. We point out that the standard security settings
on MS Windows operating systems may prevent the operation of the
program. This problem can be however easily solved by suitably
modifying the Java security options through the Windows Control
Panel. The free download of the package or of the source files
from the site sourceforge.net is also possible.

The program accepts as an input the ion concentrations at the
boundaries of the junctions. The number $n$ of ion species that
one can consider is free. For each of them one has to assign the
electrical charge and the mobility. The former is a multiple of
the elementary charge $e$, and is therefore expressed as a
relative integer $z$. The latter is defined in general as the
ratio $\mu= v/F$ between a drift velocity and the applied force
which generates the drift, and is therefore expressed in units m
s$^{-1}$ N$^{-1}$. For all the most common ions these parameters
are automatically provided by the software as soon as the ion's
name is entered in the graphical panel. In general it is however
also possible to introduce the values by hand. The final output is
represented by the junction potential $V$ and by the ionic fluxes,
i.e.~the number of ions of each species that flow through each
section of the junction per unit time.

The calculations are performed by numerically integrating the
stationary Nernst--Planck equations \cite{planck,macinnes} in
conditions of assumed local electrical quasi-neutrality at all
points of the junction. This represents an additional advantage
with respect to the available commercial software, which makes
instead use of the less accurate Henderson's equation
\cite{henderson,macinnes}. Furthermore, our program also gives the
possibility to deal with nonideal junctions, i.e.~with junctions
in which the activity of the ions cannot be identified with the
concentration. This is normally the case whenever the
concentrations are not so low. Denoting with an index $i$ the ion
species present in the junction, $1\leq i\leq n$, it is possible
to specify in an almost arbitrary way the functional dependence of
the activity $a_i$ of an ion on its concentration $c_i$, assuming
that the dependence on the concentrations of the other ions can be
neglected. This is done by giving as an input the analytical form
of the function $d\ln a_i/d\ln c_i$, which is used in the
calculation of the junction potential as explained below.

\section{Theory}

The Nernst--Planck equation describes the concentration profile of
an ion which moves inside the liquid junction under the influence
of diffusion and of the electric field. In a stationary regime and
in the absence of chemical reactions, owing to the continuity
equation the flux $\Phi_i$ of the ion is constant along the
junction. If $L$ is the length of the junction and $x$ the spatial
coordinate, with $0\leq x\leq L$, the Nernst--Planck equation for
the $x$-dependent concentration $c_i$ is
\begin{equation}\label{npeq}
\Phi_i =-D_i \frac{dc_i}{dx} +z_i\mu_i c_i eE
\end{equation}
where $D_i$ is the diffusion coefficient of the ion, while $e$,
$z_i$ and $\mu_i$ are defined as above. If we suppose that the
condition of electrical neutrality
\begin{equation}\label{neutrality}
\sum_{i=1}^n z_i c_i=0
\end{equation}
is satisfied at all points of the junction, then using (\ref{npeq})
it is possible to express the electric field as
\begin{equation}\label{eq_el}
E=\frac{A}{e\sum_{j=1}^n c_j z_j^2 \mu_j/D_j}
\end{equation}
where $A=\sum_{k=1}^n z_k \Phi_k/D_k$. Taking into account that,
as a consequence of (\ref{neutrality}), the concentration $c_n$ of
the last ion can be expressed as a function of the others as $c_n=
-\sum_{i=1}^{n-1}c_i z_i/z_n$, one finally obtains a system of
$n-1$ differential equations in the unknowns $c_1, \dots,
c_{n-1}$:
\begin{align}\label{system}
\frac{dc_i}{dx} &= \frac 1{D_i} \left(-\Phi_i+ e z_i\mu_i c_i E
\right)\nonumber \\
&=\frac 1{D_i} \left(-\Phi_i+ \frac{z_i\mu_i c_i A}
{\sum_{j=1}^{n-1} \left(z_j \mu_j/D_j-z_n \mu_n/D_n\right)c_j z_j}
\right)
\end{align}

In the hypothesis mentioned above, that the activity $a_i$ depends
only on $c_i$, the diffusion coefficients can be expressed as
\begin{equation}\label{activity}
D_i= kT\mu_i\frac{d\ln a_i}{d\ln c_i}
\end{equation}
where $k$ is Boltzmann constant and $T$ the absolute temperature.
The program calculates $D_i$ as a function of $c_i$, by using in
the right-hand side of the above equation the analytical
expression provided by the user. By default, the program assumes
that the ideal relationship $a_i=c_i$ holds. Note that in such a
case, the well-known Einstein's relation $D_i= kT\mu_i$ is
obtained.

After the user has entered the values $c_1^A, c_2^A, \dots,
c_{n-1}^A$, $c_1^B, c_2^B, \dots,
c_{n-1}^B$, of the ion concentrations at the two sides A and B
of the junction ($c_n^A$ and $c_n^B$ being fixed by the
condition of electrical neutrality),
the program integrates the differential equations (\ref{system})
for an initial guess
of the parameters $\Phi_1, \dots, \Phi_n$, taking the
concentrations on side A as initial data. Since it is assumed
that the total electrical current through the junction vanishes,
these parameters are subject to the condition
$$
\sum_{i=1}^n z_i \Phi_i=0
$$
As independent variable for the integration, the concentration
$c_1$ of the first ion is taken in place of the coordinate $x$,
which does not appear on the right side of (\ref{system}). The
procedure is automatically iterated for suitably modified values
of the fluxes $\Phi_i$, until the concentration $c_2, \dots,
c_{n-2}$ for $c_1=c_1^B$ are close to the assigned values $c^B_2,
\dots, c^B_{n-2}$ within an acceptable tolerance. Then, by using
the equations (\ref{system}) again, $x$ is in turn calculated as a
function of $c_1$. In this way the program obtains a value for the
length $L$ of the junction and for the potential
$$
V=V_B-V_A=-\int_0^L E\,dx= \int_{c_1^A}^{c_1^B} \frac {AD_1 dc_1}
{e\Phi_1 \sum_{j=1}^n c_j z_j^2 \mu_j/D_j -ez_1 \mu_1 c_1 A}
$$

The value of $L$ obtained in this way has actually no physical
meaning. The fluxes $\Phi_1, \dots, \Phi_n$, for which the
boundary conditions on side B are satisfied, are in fact not
univocally determined: their multiplication by an arbitrary
numerical factor $\lambda$ does not change either the
concentrations or the junction potential, but only amounts to
dividing by $\lambda$ the length $L$ of the junction. It follows
that only the values of $L\Phi_1, \dots, L\Phi_n$, which are
provided as the program's output together with $V$, are
independent of the arbitrary factor $\lambda$, and indeed
represent the ion fluxes multiplied by the length of the junction.
The actual length $L$ of the physical junction does not affect the
value of the potential, and is not required as an input by the
program.

In principle the results should be independent of the choice of
the ion whose concentration is taken as independent variable. The
user can verify this by repeating the calculations with different
choices, and then comparing the results. It is obvious that, in
order to suitably play the role of independent variable, the
concentration of an ion must be a monotonic function of the
position. To verify that this condition is fulfilled, it is
possible to inspect the concentration profiles along the junction,
which can be also obtained as outputs from the program.

\section{An example}

\begin{figure}[h]
\includegraphics[width=10cm]{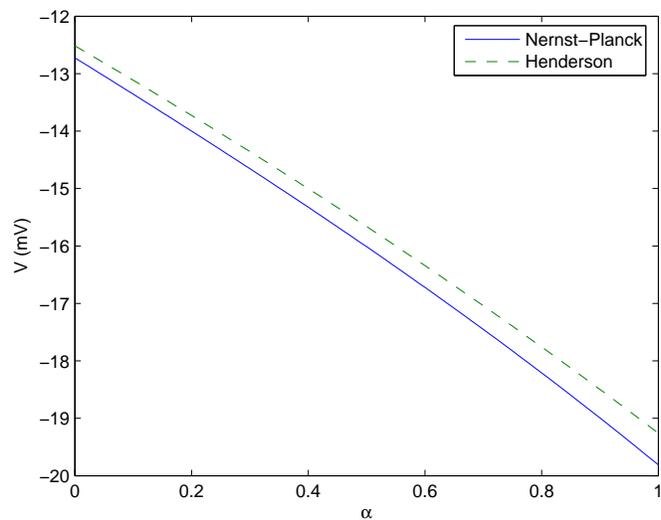}
\caption{Liquid junction potential between a solution 0.1 M of a
mixture of KOAc and KCl on one side, and a solution 0.1 M of
ZnCl$_2$ on the other. The abscissa $\alpha$ is the fraction of
acetate on the first side.}\label{fig_numeric}
\end{figure}

We have applied our program to the case in which a solution 0.1 M
of zinc chloride ZnCl$_2$ is present on side B of the junction,
while on side A there is a solution 0.1 M containing a mixture of
potassium acetate KOAc and potassium chloride KCl. If we call
$\alpha$ the fraction of acetate with respect to the total solute
on side A of the junction, for $\alpha=0$ we have only KCl, while
for $\alpha=1$ we have only KOAc. The junction is supposed to be
ideal. In Fig.~\ref{fig_numeric} we report, as a function of
$\alpha$, the liquid junction potential calculated either with our
numerical software based on the Nernst--Planck equation, or
according to the Henderson formula. For $\alpha$ approaching 0 the
numerical result converges to the value $V_0= 12.7$ mV, which can
also be obtained by analytical methods \cite{marino}, since for
$\alpha=0$ only three ion species are present in the junction. The
accordance with exact analytical solutions of the Nernst--Planck
equations has also been verified for junctions with only two ionic
species. These facts confirm the correctness of the numerical
program. We see on the other hand that, for all values of
$\alpha$, the junction potential is somewhat underestimated by the
Henderson formula.

\end{document}